\documentstyle[twocolumn,aps,prl,floats,epsf,multirow]{revtex}

\begin{document}

\draft

\wideabs{

\title{Measurement of Differences Between $J/\psi$ and $\psi'$ Suppression in p-A Collisions}

\author{
M.J.~Leitch$^f$, 
W.M.~Lee$^d$, 
M.E.~Beddo$^g$, 
C.N.~Brown$^c$,
T.A.~Carey$^f$, 
T.H.~Chang$^g$\cite{byline1},
W.E.~Cooper$^c$,
C.A.~Gagliardi$^i$,
G.T.~Garvey$^f$,
D.F.~Geesaman$^b$, 
E.A.~Hawker$^{i,f}$, 
X.C.~He$^d$,
L.D.~Isenhower$^a$,
D.M.~Kaplan$^e$,
S.B.~Kaufman$^b$,
D.D.~Koetke$^j$, 
P.L.~McGaughey$^f$, 
J.M.~Moss$^f$,
B.A.~Mueller$^b$,
V.~Papavassiliou$^g$, 
J.C.~Peng$^f$,
G.~Petitt$^d$, 
P.E.~Reimer$^{f,b}$,
M.E.~Sadler$^a$,
W.E.~Sondheim$^f$, 
P.W.~Stankus$^h$, 
R.S.~Towell$^{a,f}$,
R.E.~Tribble$^i$,
M.A.~Vasiliev$^i$\cite{byline2}, 
J.C.~Webb$^g$, 
J.L.~Willis$^a$,
G.R.~Young$^h$\\ \vspace*{9pt}
(FNAL E866/NuSea Collaboration)\\ \vspace*{9pt}
}
\address{
$^a$Abilene Christian University, Abilene, TX 79699\\
$^b$Argonne National Laboratory, Argonne, IL 60439\\
$^c$Fermi National Accelerator Laboratory, Batavia, IL 60510\\
$^d$Georgia State University, Atlanta, GA 30303\\
$^e$Illinois Institute of Technology, Chicago, IL  60616\\
$^f$Los Alamos National Laboratory, Los Alamos, NM 87545\\
$^g$New Mexico State University, Las Cruces, NM, 88003\\
$^h$Oak Ridge National Laboratory, Oak Ridge, TN 37831\\
$^i$Texas A \& M University, College Station, TX 77843\\
$^j$Valparaiso University, Valparaiso, IN 46383
}
\date{\today}

\maketitle

\begin{abstract}
Measurements of the suppression of the yield per nucleon of $J/\psi$
and $\psi'$ production for 800~GeV/c protons incident on heavy relative
to light nuclear targets have been made with very broad coverage in $x_F$
and $p_T$.
The observed  suppression is smallest at $x_F$ values of 0.25 and below
and increases at larger values of $x_F$. It is also strongest at small $p_T$.
Substantial differences between the ${\psi}'$ and $J/\psi$ are observed for
the first time in p-A collisions. The suppression for the ${\psi}'$ is stronger
than that for the $J/\psi$ for $x_F$ near zero, but becomes comparable to that
for the $J/\psi$ for $x_F > 0.6$.
\end{abstract}
\pacs{24.85.+p; 13.85.Qk; 14.40.Lb; 14.65.Dw}

} 


Strong suppression of the yield per nucleon of heavy vector mesons produced in heavy
nuclei relative to that in light nuclei has been observed in proton and pion-nucleus
collisions~\cite{e772jpsi,e789dmp,e789negjpsi,na3,pijpsi,na38}.
Similar effects have also been observed in heavy-ion collisions~\cite{hisuppr}.
This suppression exhibits strong kinematic dependences,
especially with Feynman-$x$ ($x_F$) and transverse momentum ($p_T$) of
the produced vector meson. Since the suppression of heavy vector meson
production in heavy-ion collisions is predicted to
be an important signature for the formation of the quark-gluon plasma (QGP),
it is important to understand the mechanisms that can produce similar effects
in the absence of a QGP. These mechanisms can be studied in proton-nucleus production
of vector mesons where no QGP is presumed to occur. 
Many effects have been considered\cite{vogt,aichelin,gerland,boris}
in attempting to describe the observed proton-induced charmonium
yields from nuclear targets, e.g. absorption, parton energy loss,
shadowing and feed-down from higher mass resonances, but it is clear
that no adequate understanding of the problem has been achieved. Even the
absolute cross sections are poorly understood due to poor knowledge of the
production mechanism, and most models ignore or use naive pictures of the
space-time evolution of the $c\bar{c}$ pair.
Recognizing that the production and suppression mechanisms can be identified
by their strong kinematic dependences, it is crucial to have new data with
broad kinematic coverage to challenge comprehensive descriptions of
charmonium production in nuclei.

Here we report new high statistics measurements made in Fermilab E866/NuSea of the nuclear dependence
of $J/\psi$ and $\psi'$ production for proton-nucleus collisions on Be, Fe, and
W targets. Over $3 \times 10^6$ $J/\psi$'s and
$10^5$ ${\psi}'$'s with $x_F$ between $-$0.10 and 0.93 and $p_T$
up to 4~GeV/c were observed. Previous measurements in E772~\cite{e772jpsi}
and E789~\cite{e789negjpsi,e789dmp} have suffered from limited $p_T$ acceptance and limited
statistics at larger values of $x_F$, both of which are greatly extended in these new data.

\begin{figure}[thb]
  \begin{center}
    \mbox{\epsfxsize=3in\epsffile{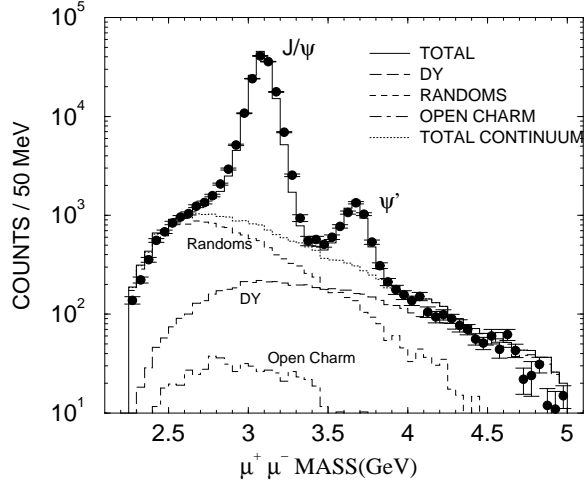}}
    \vspace*{-9pt}
  \end{center}
  \caption{Fit to the mass spectrum for the Be target in the $x_F$ range from
  0.00 to 0.05. Components in the fit are the $J/\psi$, the $\psi'$,
  Drell-Yan (long-dashed), randoms (short-dashed), and open charm (dot-dash).
  The solid curve represents the total of all fitted lineshapes and the
  dotted curve represents the continuum which is the sum of the Drell-Yan,
  randoms and open charm.}
  \vspace*{-14pt}
  \label{fig:fitbe4}
\end{figure}

E866/NuSea used a 3-dipole magnet pair spectrometer employed in
previous experiments (E605\cite{e605}, E772, and E789), modified by the
addition of new drift chambers and hodoscopes with larger
acceptance at the first tracking station and a new trigger
system~\cite{trigger}. This spectrometer was also used for other
measurements in E866/NuSea~\cite{e866dyadep,e866prlhi}.
An 800~GeV/c extracted
proton beam of up to $6 \times 10^{11}$ protons per 20 s spill
bombarded the targets used in these measurements.
A rotating wheel which was located upstream of either the first or second magnet
held thin solid targets of Be, Fe and W with thicknesses corresponding
to between 3\% and 19\% of an interaction length. After passing through
the target, the remaining beam was absorbed in a copper beam dump located
inside the large second magnet. Following the beam dump was a 13.4
interaction length absorber wall which filled the entire
aperture of the magnet, eliminated hadrons, and assured that
only muons traversed the spectrometer's detectors. These muons
were then tracked through a series of detector stations composed
of drift chambers, hodoscopes and proportional tubes. Because of improvements in
the trigger system, the coverage in $p_T$ was much broader than
in previous experiments with this spectrometer (e.g. E772), extending
to over 4~GeV/c. Beam intensity was monitored using secondary-emission
detectors.

Three magnetic field and target location configurations were used to span the full range in
$x_F$: small-$x_F$ (SXF, $-0.1 \le x_F \le 0.3$), intermediate--$x_F$ (IXF, $0.2 \le x_F \le 0.6$)
and large-$x_F$ (LXF, $0.3 \le x_F \le 0.93$).
Detailed Monte Carlo simulations of the $J/\psi$ and $\psi'$ peaks and
of the Drell-Yan continuum were used to generate lineshapes
in each bin in $x_F$ or in $p_T$. For the Drell-Yan calculations we use
MRST\cite{mrst} NLO with EKS98\cite{eskola} shadowing corrections.
The contribution to the continuum from semi-leptonic decay to muons of
open charm pairs was estimated using PYTHIA\cite{pythia} and a small correction,
less than half the statistical uncertainties,
was made for it in the SXF data set; but for the larger $x_F$ data sets
it is negligible and no corrections were made.
In addition, a detailed construction of random muon pairs using single-muon
events (which also provided a good fit to the like-sign muon mass spectra)
was used to account for the smooth random background underneath the
peaks. A maximum-likelihood method was used for fitting that took into account
the statistical uncertainty of the data and of the Monte Carlo and randoms\cite{hmcmll}.
Figure 1 shows a typical fit to a mass spectrum using these components.
Since the rates in the various detectors were nearly equal
for the different targets, a correction for rate-dependent inefficiencies
was not necessary.

We present our results in terms of $\alpha$, where
$\alpha$ is obtained by assuming the cross section dependence on nuclear
mass, $A$, to be of the form $ \sigma_A = \sigma_N \times A^{\alpha}$,
where $\sigma_N$ is the cross section on a nucleon.
For the SXF data,
$\alpha$ was obtained using Be and two different thickness W targets, while for the IXF and
LXF data, Be, Fe and W targets were used. The SXF data from the two W targets verified
that no corrections for secondary production were necessary.
The $p_T$ dependence of $\alpha$ is shown in Fig.~2, where we see essentially
the same increase in $\alpha$ for all $x_F$ ranges for both
the $J/\psi$ and the $\psi'$, as well as for the 200~GeV/c NA3 data\cite{na3}.
This increase is characteristic of multiple scattering of the incident
parton and of the nascent $c\bar{c}$ in the final state.
Note that for the IXF data the $p_T$ acceptance is truncated at about 2~GeV/c
because a more restrictive trigger was used.

\begin{figure}[thb]
  \begin{center}
    \vspace*{-36pt}
    \mbox{\epsfxsize=3in\epsffile{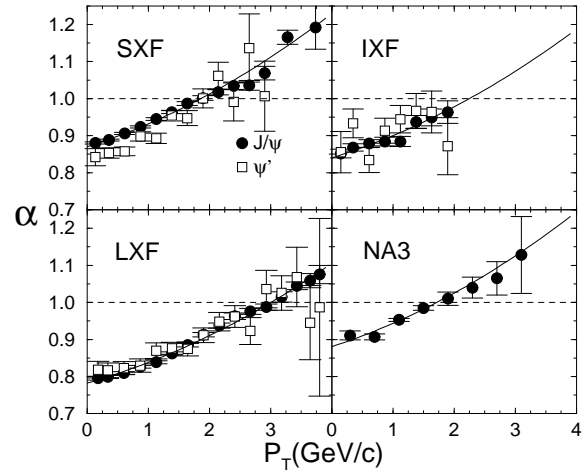}}
    \vspace*{-9pt}
  \end{center}
  \caption{$\alpha$ versus $p_T$ for $J/\psi$ (solid circles) and
  $\psi'$ (open boxes) production by 800~GeV/c protons.
  Results are shown for the three data sets -- SXF, IXF and LXF (see text) -- 
  which have $\langle x_F \rangle = $ 0.055, 0.308 and 0.480, respectively.
  Only statistical uncertainties are shown. An additional systematic uncertainty of 0.5\%
  is not included.
  Also shown are the NA3 results at 200~GeV/c whose $x_F$ range can be seen in Fig.~4.
  The solid curves represent the parameterization discussed in the text.}
  \label{fig:apt4}
  \vspace*{-4pt}
\end{figure}

Previous experiments such as E772 have had a limited acceptance in $p_T$
which varied with $x_F$. Since the value of $\alpha$ depends strongly on
$p_T$ this can cause a distortion of the apparent shape of $\alpha$ versus
$x_F$. The improvements in the E866/NuSea trigger allowed a much broader
$p_T$ acceptance than in these earlier measurements. However, for the lowest
values of $x_F$ at each spectrometer setting our $p_T$ acceptance still
becomes somewhat restricted. For the results presented here we have corrected the
values of $\alpha(x_F)$ using a detailed simulation of our acceptance and
a differential cross section shape versus $p_T$ derived from our data.

\begin{figure}[thb]
  \begin{center}
    \vspace*{-7pt}
    \mbox{\epsfxsize=3in\epsffile{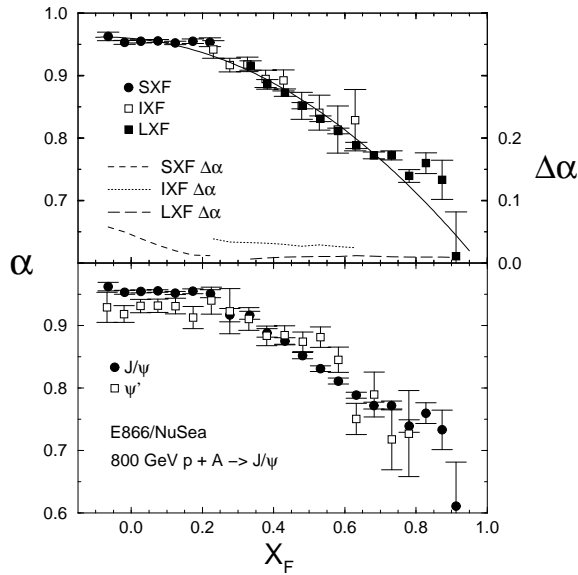}}
    \vspace*{-10pt}
  \end{center}
  \caption{$\alpha$ for the $J/\psi$ versus $x_F$ for the three different data
  sets (top) and for the $J/\psi$ and $\psi'$ after the data sets are combined (bottom).
  Values are corrected for the $p_T$ acceptance, as discussed in the text.
  These corrections ($\Delta\alpha$) have a maximum value of 0.06 and are shown
  using the right-hand vertical scale in the top panel. The relative systematic
  uncertainty between $\alpha$ for the $J/\psi$ and $\psi'$ is estimated to
  be 0.003, while the absolute systematic uncertainty is 0.01 in $\alpha$;
  neither is included here.
  The solid curve represents the parameterization discussed in the text.}
  \label{fig:axf2}
  \vspace*{-10pt}
\end{figure}

The resulting dependence of $\alpha$ on $x_F$ is shown in Fig.~3
and listed in Table I.
The systematic uncertainty of $1\%$ in the corrected $\alpha$ is dominated
by the $p_T$ acceptance correction. $\alpha$ for the $J/\psi$ is largest at values of
$x_F$ of 0.25 and below but strongly decreases at
larger values of $x_F$.
For the ${\psi}'$ $\alpha$ is smaller than for the $J/\psi$ for
$x_F < 0.2$, remains relatively constant up to
$x_F$ of 0.5 (becoming slightly larger than for the $J/\psi$) and
then falls to values consistent with those for the $J/\psi$ for $x_F > 0.6$.
The significance of the overall $J/\psi$, $\psi'$ difference for $x_F < 0.2$ is
about 4 sigma with respect to the statistical and relative systematic
uncertainties. This difference is consistent with less accurate results
obtained by NA38 for p-A at 450 GeV/c\cite{na38}, but is inconsistent with
the quoted NA38 result that also included the p-p and p-d data from NA51.
Although slightly larger $\alpha$ values for the $\psi$' than for the $J/\psi$
can be seen near $x_F = 0.55$, we should point out that if instead we emphasize
the velocity of the $c{\bar c}$ and plot $\alpha$ versus rapidity, then the
agreement is quite good in this region.
The reduced $\alpha$ at small $x_F$ is also evident in Fig.~2
where $\alpha$ for the $\psi'$ falls consistently below that for the $J/\psi$ at
low $p_T$ for the SXF data set.

Our results for the $J/\psi$ $\alpha$ can be represented for convenience by the
simple parameterizations shown as solid lines in Figs.~2 and 3:
$\alpha(x_F) = 0.960 (1 - 0.0519 x_F - 0.338 x_F^2)$,
and $\alpha(p_T) = A_i (1 + 0.0604 p_T + 0.0107 {p_T}^2)$,
where $A_i = 0.870$, $0.840$, $0.782$ and $0.881$ for the SXF, IXF and LXF datasets and
for the NA3 data, respectively.

\begin{table}[tb]

  \caption{$\alpha$ versus $x_F$\protect\cite{xfdef} for the $J/\psi$ and $\psi'$.
           $\alpha$ is defined by $\sigma_A = \sigma_N \times A^{\alpha}$
           and is equal to one if there is no suppression and the cross section
           scales simply as the number of nucleons. The average momentum fraction
           of the struck parton, $x_2$\protect\cite{x2def}, and the center-of-mass
           rapidity, $y_{cm}$, are also shown.
           An additional systematic uncertainty of 1\% is not included here.}

  \label{tab:1}
  \begin{tabular}{ccccccc}
 $\langle x_F\rangle$ & $\langle y_{cm}\rangle_{J/\psi}$
 & $\langle x_2\rangle_{J/\psi}$
 & $\alpha_{J/\psi}$
 &$\langle y_{cm}\rangle_{\psi'}$ & $\langle x_2\rangle_{\psi'}$
 & $\alpha_{\psi'}$\\ \hline
  $-$0.065 &$-$0.390 &0.1192 &0.962(7)    &$-$0.344 &0.1346 &0.929(24) \\
  $-$0.019 &$-$0.115 &0.0902 &0.953(3)    &$-$0.104 &0.1056 &0.918(14) \\
   0.027 &0.161 &0.0679 &0.955(2)    &0.132 &0.0828 &0.931(11) \\
   0.075 &0.433 &0.0511 &0.955(2)    &0.369 &0.0645 &0.932(11) \\
   0.124 &0.680 &0.0395 &0.952(3)    &0.588 &0.0513 &0.931(12) \\
   0.173 &0.896 &0.0316 &0.955(4)    &0.785 &0.0418 &0.913(18) \\
   0.223 &1.091 &0.0262 &0.951(6)    &0.974 &0.0347 &0.940(22) \\
   0.277 &1.288 &0.0213 &0.917(11)   &1.144 &0.0293 &0.923(36) \\
   0.332 &1.427 &0.0182 &0.916(6)   &1.281 &0.0253 &0.910(18) \\
   0.381 &1.551 &0.0160 &0.888(7)   &1.401 &0.0223 &0.884(16) \\
   0.431 &1.663 &0.0142 &0.875(6)    &1.512 &0.0199 &0.885(15) \\
   0.481 &1.764 &0.0128 &0.852(5)    &1.614 &0.0179 &0.874(16) \\
   0.531 &1.858 &0.0117 &0.831(5)    &1.705 &0.0163 &0.881(16) \\
   0.582 &1.945 &0.0107 &0.811(5)    &1.791 &0.0150 &0.845(20) \\
   0.632 &2.026 &0.00984 &0.789(6)   &1.869 &0.0138 &0.751(25) \\
   0.682 &2.098 &0.00916 &0.772(5)   &1.942 &0.0129 &0.790(36) \\
   0.732 &2.166 &0.00855 &0.772(7)   &2.009 &0.0120 &0.718(49) \\
   0.781 &2.228 &0.00804 &0.739(10)  &2.071 &0.0113 &0.727(69) \\
   0.828 &2.286 &0.00760 &0.760(17)  & \\
   0.873 &2.338 &0.00723 &0.733(32)  & \\
   0.913 &2.383 &0.00698 &0.611(71)  & \\
  \end{tabular}
  \vspace*{-20pt}
\end{table}

A comparison of our results with earlier data from E772
at 800~GeV/c\cite{e772jpsi} and also with NA3
at 200~GeV/c\cite{na3} is shown in Fig. 4. It illustrates that
the suppression seen for $J/\psi$ production scales with $x_F$
but not with $p^{LAB}_{J/\psi}$ above about 90~GeV/c,
which corresponds to $x_F > 0.05$ for our data and to $x_F > 0.4$ for NA3.
Also of interest in these figures is a comparison of our results
with those of E772. At the small-$x_F$ end of the E772 data their published
results drop significantly below our results. As was discussed above, the
E772 data have severe narrowing of the $p_T$ acceptance for their smallest
$x_F$ bins, and a large correction that could easily bring the E772 points
into agreement with our data is expected. Similar arguments hold for the
E789 $J/\psi$ data (not shown) at small to negative $x_F$\cite{e789negjpsi},
where we estimate
about an 8\% correction which would bring those results into agreement with
ours. On the other hand, the large $x_F$ results from E789\cite{e789dmp}
appear to be high by more than their systematic uncertainty of 2.5\%.

\begin{figure}[thb]
  \begin{center}
    \mbox{\epsfxsize=3in\epsffile{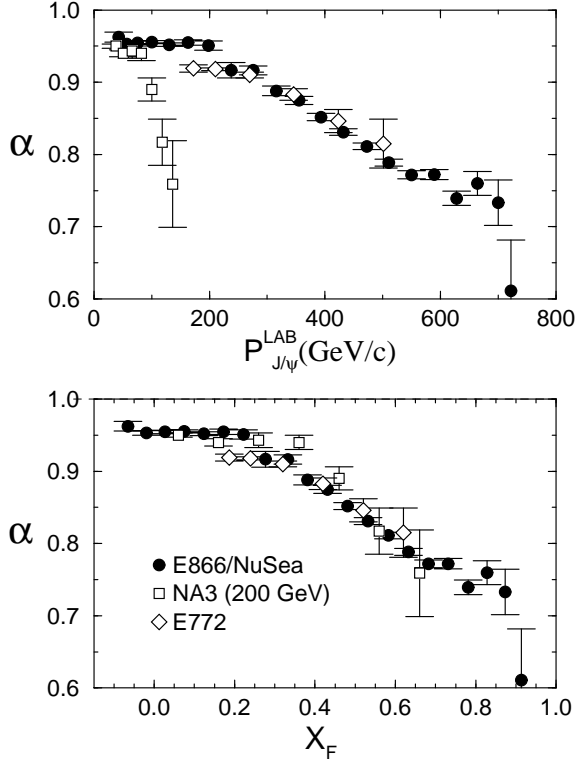}}
  \end{center}
  \vspace*{-10pt}
  \caption{$\alpha$ versus $x_F$ and versus $p^{LAB}_{J/\psi}$ for the $J/\psi$ from E866/NuSea (800~GeV/c) (solid circles)
  compared to E772 (open diamonds) and NA3 (200~GeV/c) (open squares) showing the scaling with
  $x_F$ (bottom) and lack of scaling with $p^{LAB}_{J/\psi}$ (top).}
  \label{fig:axfplab}
  \vspace*{-15pt}
\end{figure}

The suppression of $J/\psi$ production near $x_F=0$ is usually thought to
be caused by absorption, the dissociation of the $c\bar{c}$ pair by
interactions with the nucleus or comovers \cite{vogt} into separate quarks
that eventually hadronize into $D$ mesons.  This model is supported by
both the increased suppression of the $\psi'$ that we observe near
$x_F=0$ and the absence of suppression of $D$ meson production in the
same kinematic region \cite{e789d}.  At small $x_F$, the velocity of the
$c\bar{c}$ pair is low enough that it may hadronize within the nucleus, so
the larger $\psi'$ would be absorbed more strongly \cite{aichelin,gerland}.  However,
the observed constancy of $\alpha$ for both the $J/\psi$ and the $\psi'$ up to
$y_{cm} \simeq 1$ complicates this interpretation since these models predict that
faster $c\bar{c}$ pairs above $x_F \simeq 0.1$ would experience similar
absorption, whether they eventually hadronize outside the nucleus into a
$J/\psi$ or a $\psi'$. At larger values of $x_F$, above $0.3$, our data does show
similar suppression for the $J/\psi$ and $\psi'$.
Furthermore, if absorption by the nuclear
medium is the dominant suppression mechanism, the effect should scale
with $p^{LAB}_{J/\psi}$, but Fig.~4 shows that scaling breaks
down in the middle of the region where we observe $\alpha$ to be constant.

Shadowing of the small-$x$ target gluon distributions is also thought to play a role
in the observed suppression, but current estimates\cite{vogt,eskola}
predict at most a few percent drop in $\alpha(x_F)$, even at the largest $x_F$ values.
Also, as is seen for our data (but not shown) and was seen previously\cite{e772jpsi},
there is a lack of scaling with $x_2$, which is related to that shown above for
$p^{LAB}_{J/\psi}$ since $x_2 \propto 1/p^{LAB}_{J/\psi}$.
This appears to rule out large contributions from shadowing. 
Our studies\cite{e866dyadep}
show that for Drell-Yan the dominant nuclear effect is shadowing
of the anti-quark distributions and that the energy-loss of the incident
quark is small. Although the incoming gluon's energy loss is expected to be larger by a
color factor of $9/4$ and the additional energy loss of the outgoing $c{\bar{c}}$
may be as large as that of a gluon, we still expect the overall
contribution of energy loss to be small for resonance production.

In conclusion, we have presented new data for the suppression of $J/\psi$
and $\psi'$ production in heavy versus light nuclei for 800~GeV/c proton-nucleus
collisions. The kinematic coverage in $x_F$ ($-$0.10 to 0.93) and $p_T$ (0 - 4~GeV/c)
and statistical accuracy surpass that of previous experiments. Corrections
are made to the data to account for the narrowing $p_T$ acceptance at the
smaller values of $x_F$.
The largest value of $\alpha$ (integrated over $p_T$) of about 0.95 is seen
at $x_F$ near 0.25 and below with strongly falling values for larger $x_F$.
The most striking new result is that the suppression for the ${\psi}'$ is
stronger than that for the $J/\psi$ at $x_F$ near zero.

We thank Ramona Vogt and Boris Kopeliovich for many useful discussions 
and the Fermilab Particle Physics, Beams and
Computing Divisions for their assistance.
This work was supported in part by the U.S. Department of
Energy.


\begin{references}
\bibitem{xfdef} We calculate $x_F = p_L^{CM}/(\sqrt{s}/2)$, neglecting a small ($1-\tau \simeq 0.6\%$) correction
  for the $J/\psi$ mass.
\bibitem{x2def} We calculate $x_2$ by ignoring the soft gluon that accompanies
  the $c\bar{c}$ pair in the final state in the same way one would calculate $x_2$
  for the leading-order Drell-Yan process.
\end{references}
\end{document}